\documentclass{appolb}
\usepackage{graphicx}
\usepackage{epsfig}
\usepackage{epsf}
\usepackage{graphics}
\usepackage{amsmath}
\newcommand \vt[1]{\bf #1}

\begin{document}
\title{$CP$ INVARIANCE STUDY OF $J/\psi\to\Lambda\bar\Lambda$ AND $\Lambda$
 NONLEPTONIC DECAYS IN HELICITY FRAME%
}
\author{Bin ZHONG
\address{Department of Physics and Institute of Theoretical Physics,\\
Nanjing Normal University, Nanjing, Jiangsu 210023, P. R. China}
\\
\vspace{5mm}
Guangrui LIAO
\address{College of physical science and technology,\\
Guangxi Normal university, Guilin, Guangxi 541004, P. R. China}
}
\maketitle
\begin{abstract}
We present the joint helicity amplitudes
for $J/\psi \to \Lambda \bar{\Lambda}$, $\Lambda(\bar\Lambda)$
decays to different final states in the helicity frame.
Two observables to search for $CP$ violation in
$J/\psi\to\Lambda\bar\Lambda$ can be expressed with the
information of helicity angles of baryon and antibaryon.
Four decay parameters of $\Lambda$ and $\bar\Lambda$, namely,
$\alpha_-,\alpha_+,\alpha_0$ and $\bar\alpha_0$, can be obtained with the joint
helicity amplitude equations by the likelihood fit method.
With the data sample of $10^{10}$
$J/\psi$ decays accumulated by BESIII,
the precision of the measurements is estimated to be about $10^{-3}$.
\end{abstract}

\section{Introduction}
$\Lambda$$\bar\Lambda$ decay is one of the octet baryon-antibaryon pairs
decays of $J/\psi$ and other charmonium states. The branching ratio of $J/\psi\to\Lambda\bar\Lambda$ has
been measured by BES\cite{bes} and CLEO\cite{cleoc} collaboration. The world average value is
$(1.61\pm0.15)\times10^{-3}$\cite{pdg}.
It is noted that this decay channel is very special to study the $CP$ invariance
not only in $J/\psi\to\Lambda\bar\Lambda$ but also in the nonleptonic decay of $\Lambda(\bar\Lambda)$.

$CP$ violation in $J/\psi\to\Lambda\bar\Lambda$ decay is studied in Ref.\cite{jpsicp},
Two observables are suggested
to test $CP$ invariance,
\begin{eqnarray}\label{jpsiA}
A_{J/\psi}=\theta(\hat{\bf p}\cdot(\hat{\bf q}_{\bf 1}\times\hat{\bf q}_{\bf 2})) - \theta(-\hat{\bf p}\cdot(\hat{\bf q}_{\bf 1}\times\hat{\bf q}_{\bf 2})) ,
\end{eqnarray}

\begin{eqnarray}\label{jpsiB}
B=\hat{\bf p}\cdot(\hat{\bf q}_{\bf 1}\times\hat{\bf q}_{\bf 2}),
\end{eqnarray}
where $\theta(x)$ is 1 if $x>0$ and is zero if $x<0$.
$\hat{\bf p}$, $\hat{\bf q}_1$ and $\hat{\bf q}_2$ are the momentum unit vectors of
$\Lambda$, proton and anti-proton. Any nonzero values for them signal $CP$ violation.
Beside $J/\psi\to\Lambda\bar\Lambda$, the measurement can be carried in other charmonium states decay to $\Lambda\bar\Lambda$ experimentally.

$CP$ violation can also be studied in nonleptonic decays of $\Lambda$.
Nonleptonic hyperon decays have long been known as an ideal
laboratory to study the parity violation\cite{Lee&yang}. Considering
a nonleptonic decay of the hyperon $Y$, the angular
distribution of the baryon in the center-of-mass (CM) system of $Y$
takes the form
${d N\over d\Omega}\propto 1+ \alpha_Y{\bf P} \cdot \hat{\vt p}$,
where $\bf P$ is the polarization vector of the hyperon,
$\hat{\vt{p}}$
is the momentum unit vector of the baryon, and $\alpha_Y$ is
the hyperon decay parameter,
which characterizes the parity violation in the decays.
Taking $\Lambda\to p\pi^-$ as an example, a $CP$-odd observable, $A_{\Lambda}$,
can be defined as
\begin{eqnarray}\label{lambdacp}
A_{\Lambda}=\frac{\alpha_-+\alpha_+}{\alpha_--\alpha_+},
\end{eqnarray}
where $\alpha_-$ is the decay parameter of $\Lambda\to p\pi^-$,
$\alpha_+$ is the decay parameter of $\bar\Lambda\to\bar p\pi^+$.
If $\Lambda$ decays to $n\pi^0$ and $\bar\Lambda$ decays to $\bar n\pi^0$,
$\alpha_-$ and $\alpha_+$ should be replaced by $\alpha_0$ and $\bar\alpha_0$.
If $CP$ is conserved, this observable vanishes for $\alpha_-=-\alpha_+$
or $\alpha_0=-\bar\alpha_0$.
Any nonzero value implies evidence for $CP$ asymmetry in $\Lambda$
decays.
This asymmetry has been previously performed at
$p\bar p$ colliders by the R608\cite{6r608} and PS185\cite{7ps185} collaborations,
and at an $e^+e^-$ collider by DM2\cite{8dm2} collaboration. The latest result is
measured by BES\cite{9bes} collaboration, although the precision has been improved much,
it is insufficient to observe $CP$ violation at the level predicted by the
standard model.

The precise measurement of the $\Lambda$ decay parameter also
plays an important role in the determination of $\Omega^-$ and
$\Xi^-$ decay parameters.
To note that the non-polarized
$\Omega^-$ or $\Xi^-$ decays can produce polarized
$\Lambda$ particle. Namely, $\Lambda$ is the daughter particle in
the decays of $\Omega^-$ and $\Xi^-$. In the $\Lambda$ rest frame, the
angular distribution of the proton takes
the form of ${d N\over d\cos\theta}\propto 1+\alpha_{
\Omega(\Xi)}\alpha_-\cos\theta$. Experimentally, the
extraction of $\alpha_{\Omega(\Xi)}$ from the product
$\alpha_{\Omega(\Xi)}\alpha_-$ is dependent on
the value of $\alpha_-$. The accuracy of
$\alpha_{\Omega(\Xi)}$ measurement is dependent
on the accuracy of $\alpha_-$. The situation is the same for
$\bar\Omega^+$ and $\bar\Xi^+$ which can produce polarized $\bar\Lambda$.
So, the measurement of
$\alpha_-$ and $\alpha_+$ plays a unique role in other hyperon
decay parameters measurement.

In this paper, based on the study in Refs.\cite{jpsicp}\cite{10lambda}, we detail the
information on the $CP$ observables in $J/\psi\to\Lambda\bar\Lambda$,
and the decay parameters of $\Lambda$ and $\bar\Lambda$ with helicity amplitude
analysis. The similar study is applied for $\psi(2S)\to\gamma\chi_{cJ}\to\gamma\Lambda\bar\Lambda$\cite{11liao}.
Nowadays, $10^{10}$ $J/\psi$ decays have been
accumulated by BESIII detector, the advantage of this
work is obvious and high accuracy can be achieved for the statistics.

\section{Helicity amplitudes analysis of $J/\psi\to\Lambda\bar\Lambda$,$\Lambda\to B\pi$, $\bar\Lambda\to \bar B\pi$}
\begin{figure}
\centerline{\includegraphics[width=6cm,height=4cm]{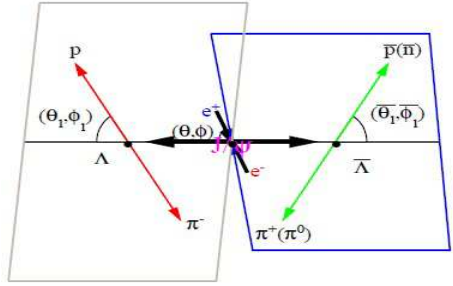}}
\caption{Definition of the helicity frame for $J/\psi\to\Lambda\bar\Lambda$, $\Lambda\to B\pi$, $\bar\Lambda\to \bar B\pi$
\label{heli}}
\end{figure}

The helicity amplitudes for $J/\psi\to\Lambda\bar\Lambda$, $\Lambda\to B\pi$, $\bar\Lambda\to \bar B\pi$
decays are constructed in the helicity frame defined as:
\begin{enumerate}
  \item In $J/\psi\to\Lambda\bar\Lambda$, the z-axis of the $J/\psi$ rest frame
  is along $\Lambda$ outgoing direction, which changes from event to event. The
  $e^+$ beam is along the direction of the solid angle ($\theta,\phi$).
  \item For $\Lambda$ decay $\Lambda\to B\pi$, the solid angle of the daughter particle
  $\Omega_1(\theta_1,\phi_1)$ is refereed to the $\Lambda$ rest frame, where the z-axis is taken along the outgoing direction of $\Lambda$ in its mother particle rest frame. The helicity frame of $\bar\Lambda$ has a similar definition which is described by the solid angle $\bar\Omega_1(\bar\theta_1,\bar\phi_1)$.
\end{enumerate}
Fig.\ref{heli} shows the definition of the helicity frame for $J/\psi\to\Lambda\bar\Lambda$, $\Lambda\to B\pi$, $\bar\Lambda\to \bar B\pi$.

The joint helicity amplitudes of $J/\psi\to\Lambda\bar\Lambda$, $\Lambda\to B\pi$,
 $\bar\Lambda\to \bar B\pi$ can be expressed by $A_{\lambda,\bar{\lambda}}$, $B_{\lambda_B}$ and $\bar{B}_{\lambda_{\bar{B}}}$ as:
\begin{eqnarray}\label{amp}
|\mathcal{M}|^2 & \propto \sum\limits_{\lambda,\bar\lambda,\lambda',\bar\lambda',\lambda_B,\bar\lambda_{\bar B}}
\rho^{\lambda-\bar\lambda,\lambda'-\bar\lambda'}(\theta,\phi)\times  \nonumber \\
& A_{\lambda,\bar\lambda}A^*_{\lambda',\bar\lambda'}B_{\lambda_B}B^*_{\lambda_B}
\bar B_{\lambda_{\bar B}}\bar B^*_{\lambda_{\bar B}}\times \nonumber \\
& D^{1/2*}_{\lambda,\lambda_B}(\Omega_1)D^{1/2}_{\lambda',\lambda_B}(\Omega_1)\times \nonumber \\
& D^{1/2*}_{\bar\lambda,\bar\lambda_B}(\bar\Omega_1)D^{1/2}_{\bar\lambda',\bar\lambda_B}(\bar\Omega_1),
\end{eqnarray}
where $\lambda,\bar{\lambda}$,$\lambda_B,\bar\lambda_B$ are
the helicity values for $\Lambda, \bar\Lambda$, baryon and anti-baryon.

$\rho^{(i,j)}(\theta,\phi) =
\sum_{k=\pm1}D^1_{i,k}(\Omega)D^{1*}_{j,k}(\Omega)$ is the
density matrix for the $J/\psi$ produced in $e^+e^-$ annihilation. The element
$\rho^{(i,j)}$ is equal to $\rho^{(j,i)\ast}$ and equal to $(-1)^{(i+j)}\rho^{(-i,-j)}$.
The density matrix elements are shown in Tab.\ref{rhomatrix}.

\begin{table}
\renewcommand\arraystretch{2.0}
\begin{center}
\caption{$J/\psi$ density matrix elements}\label{rhomatrix}
\vspace{5mm}
\begin{tabular}{cc}
\hline
$\rho^{(1,1)}(\theta,\phi)=\frac{1+\cos^2\theta}{2}$ & $\rho^{(0,0)}(\theta,\phi)=\sin^2\theta$ \\
$\rho^{(1,0)}(\theta,\phi)=\frac{\sin\theta\cos\theta}{\sqrt{2}}e^{-i\phi}$ & $\rho^{(1,-1)}(\theta,\phi)=\frac{\sin^2\theta}{2}e^{-2i\phi}$  \\
\hline
\end{tabular}
\end{center}
\end{table}

$D^J_{\lambda,\lambda'}(\Omega) \equiv
D^J_{\lambda,\lambda'}(\phi,\theta,0)$ is the Wigner D-function. The standard Wigner D-function is defined as:
\begin{eqnarray}\label{wignerD}
D^j_{m,m'}(\alpha\beta\gamma)=&\sum\limits_{k=0}^{j+m}(-1)^{m'-m+k}\frac{\sqrt{(j-m)!(j+m)!(j-m')!(j+m')!}}{k!(j+m-k)!(j-m'-k)!(k+m'-m)!}\times \nonumber \\
& e^{-im'\alpha}e^{-im\gamma}(\cos\frac{\beta}{2})^{2j+m-m'-2k}(\sin\frac{\beta}{2})^{m'-m+2k},
\end{eqnarray}

The final baryons in $\Lambda$ and $\bar\Lambda$ decays are $p, \bar p, n, \bar n$.
The possible helicity values for $p, \bar p, n, \bar n$ are $1/2$ or $-1/2$, so the
density matrix elements of $D^J_{\lambda,\lambda'}(\Omega)$ with $J=1/2$ are shown in Tab.\ref{Dmatrix}.

\begin{table}
\renewcommand\arraystretch{2.5}
\begin{center}
\caption{Density matrix elements of $D^J_{\lambda,\lambda'}(\Omega)$ with $J=1/2$}\label{Dmatrix}
\vspace{5mm}
\begin{tabular}{cc}
\hline
$D^{\frac{1}{2}}_{\frac{1}{2},\frac{1}{2}}(\Omega)=e^{-i\frac{\phi}{2}}\cos\frac{\theta}{2}$ & $D^{\frac{1}{2}}_{\frac{1}{2},-\frac{1}{2}}(\Omega)=-e^{-i\frac{\phi}{2}}\sin\frac{\theta}{2}$ \\
$D^{\frac{1}{2}}_{-\frac{1}{2},\frac{1}{2}}(\Omega)=e^{i\frac{\phi}{2}}\sin\frac{\theta}{2}$ & $D^{\frac{1}{2}}_{-\frac{1}{2},-\frac{1}{2}}(\Omega)=e^{i\frac{\phi}{2}}\cos\frac{\theta}{2}$  \\
\hline
\end{tabular}
\end{center}
\end{table}

From parity invariance, $A_{-\lambda,-\lambda'}=A_{\lambda,\lambda'}$.
so the decay parameters in $\Lambda$ and $\bar\Lambda$ decays are defined as:
\begin{eqnarray}\label{alpham}
\alpha_-=\alpha(\Lambda\to p \pi^-)=\frac{|B_{1/2}|^2-|B_{-1/2}|^2}{|B_{1/2}|^2+|B_{-1/2}|^2},
\end{eqnarray}
\begin{eqnarray}\label{alphap}
\alpha_+=\alpha_(\bar\Lambda\to \bar p \pi^+)=\frac{|\bar{B}_{1/2}|^2-|\bar{B}_{-1/2}|^2}{|\bar{B}_{1/2}|^2
+|\bar{B}_{-1/2}|^2},
\end{eqnarray}
\begin{eqnarray}\label{alpha0}
\alpha_0=\alpha(\Lambda\to n \pi^0)=\frac{|B_{1/2}|^2-|B_{-1/2}|^2}{|B_{1/2}|^2+|B_{-1/2}|^2},
\end{eqnarray}
\begin{eqnarray}\label{alpha0b}
\bar\alpha_0=\alpha_(\bar\Lambda\to \bar n \pi^0)=\frac{|\bar{B}_{1/2}|^2-|\bar{B}_{-1/2}|^2}{|\bar{B}_{1/2}|^2
+|\bar{B}_{-1/2}|^2},
\end{eqnarray}
If $CP$ invariance is conserved, one has $B_\lambda=-\bar
B_{-\lambda}$. $\alpha_- = -\alpha_+$ and $\alpha_0 = -\bar\alpha_0$ can be gotten.
The angular distribution of baryons($p,n$) and anti-baryons($\bar p,\bar n$) in $\Lambda(\bar\Lambda)$ rest frame can be written as:
\begin{eqnarray}\label{angularp}
\frac{d N}{d\Omega} \propto 1+ \alpha_-|{\bf P}|\cos\theta_1 ~~~~\Lambda\to p \pi^-,
\end{eqnarray}
\begin{eqnarray}\label{angularpbar}
\frac{d N}{d\Omega} \propto 1+ \alpha_+|{\bf P}|\cos\bar\theta_1 ~~~~\bar\Lambda\to \bar p \pi^+, \end{eqnarray}
\begin{eqnarray}\label{angularn}
\frac{d N}{d\Omega} \propto 1+ \alpha_0|{\bf P}|\cos\theta_1 ~~~~\Lambda\to n \pi^0,
\end{eqnarray}
\begin{eqnarray}\label{angularnbar}
\frac{d N}{d\Omega} \propto 1+ \bar\alpha_0|{\bf P}|\cos\bar\theta_1 ~~~~\bar\Lambda\to \bar n \pi^0,
\end{eqnarray}
The product of decay parameter and polarization of $\Lambda(\bar\Lambda)$
can be gotten by fitting the angular distribution of the baryons(anti-baryons). If the
polarization of $\Lambda(\bar\Lambda)$ could be measured, the decay parameter could be
extracted, but in $e^+e^-$ collision, the average value of polarization for the produced $\Lambda(\bar\Lambda)$ is zero, so, experimentally, the decay parameter can not be measured
in this way.

From Eq.(\ref{amp}), the angular distribution parameter of $\Lambda(\bar\Lambda)$
can be defined as:
\begin{eqnarray}\label{ang}
\alpha={|A_{1/2,-1/2}|^2-2|A_{1/2,1/2}|^2\over
|A_{1/2,-1/2}|^2+2|A_{1/2,1/2}|^2},
\end{eqnarray}

if the
normalization condition is selected as
$|A_{1/2,-1/2}|^2 + 2|A_{1/2,1/2}|^2 = 1$ , one has:
\begin{eqnarray}\label{norm}
|A_{1/2,-1/2}|^2={1+\alpha \over 2}
\textrm{~and~}|A_{1/2,1/2}|^2={1-\alpha\over 4}.
\end{eqnarray}

Combining Eqs.(\ref{amp})(\ref{alpham})(\ref{alphap})(\ref{ang})(\ref{norm}) and integrating over $\phi$,
Eq.(\ref{amp}) is simplified as:
\begin{eqnarray}\label{ampsim}
& \frac{d|\mathcal{M}|^2}{d(\cos\theta)d\Omega_1d\bar{\Omega}_1}
\propto
(1-\alpha)\sin^2\theta\times \nonumber \\
&[1+\alpha_-\alpha_+(\cos\theta_1\cos\bar\theta_1+\sin\theta_1\sin\bar\theta_1\cos(\phi_1+\bar\phi_1))] \nonumber \\
&-(1+\alpha)(1+\cos^2\theta)(\alpha_-\alpha_+\cos\theta_1\cos\bar\theta_1-1),
\end{eqnarray}
Eq.(\ref{ampsim}) is the helicity amplitude equation for $J/\psi\to\Lambda\bar\Lambda\to p\pi^- \bar p\pi^+$. With different final states of $\Lambda$ and $\bar\Lambda$ decays, according to
Eqs.(\ref{alpham})(\ref{alphap})(\ref{alpha0})(\ref{alpha0b}), one can also get helicity amplitude equations for
$J/\psi\to\Lambda\bar\Lambda$ with different final states.
\begin{itemize}
\item $J/\psi\to\Lambda\bar\Lambda\to p\pi^- \bar n\pi^0$
\begin{eqnarray}\label{alpham0bar}
& \frac{d|\mathcal{M}|^2}{d(\cos\theta)d\Omega_1d\bar{\Omega}_1}
\propto
(1-\alpha)\sin^2\theta\times \nonumber \\
&[1+\alpha_-\bar\alpha_0(\cos\theta_1\cos\bar\theta_1+\sin\theta_1\sin\bar\theta_1\cos(\phi_1+\bar\phi_1))] \nonumber \\
&-(1+\alpha)(1+\cos^2\theta)(\alpha_-\bar\alpha_0\cos\theta_1\cos\bar\theta_1-1),
\end{eqnarray}
\item $J/\psi\to\Lambda\bar\Lambda\to n\pi^0 \bar p\pi^+$
\begin{eqnarray}\label{amp0p}
& \frac{d|\mathcal{M}|^2}{d(\cos\theta)d\Omega_1d\bar{\Omega}_1}
\propto
(1-\alpha)\sin^2\theta\times \nonumber \\
&[1+\alpha_0\alpha_+(\cos\theta_1\cos\bar\theta_1+\sin\theta_1\sin\bar\theta_1\cos(\phi_1+\bar\phi_1))] \nonumber \\
&-(1+\alpha)(1+\cos^2\theta)(\alpha_0\alpha_+\cos\theta_1\cos\bar\theta_1-1),
\end{eqnarray}
\item $J/\psi\to\Lambda\bar\Lambda\to n\pi^0 \bar n\pi^0$
\begin{eqnarray}\label{amp00bar}
& \frac{d|\mathcal{M}|^2}{d(\cos\theta)d\Omega_1d\bar{\Omega}_1}
\propto
(1-\alpha)\sin^2\theta\times \nonumber \\
&[1+\alpha_0\bar\alpha_0(\cos\theta_1\cos\bar\theta_1+\sin\theta_1\sin\bar\theta_1\cos(\phi_1+\bar\phi_1))] \nonumber \\
&-(1+\alpha)(1+\cos^2\theta)(\alpha_0\bar\alpha_0\cos\theta_1\cos\bar\theta_1-1),
\end{eqnarray}

\end{itemize}

\section{$CP$ violation of $J/\psi\to\Lambda\bar\Lambda$, $\Lambda\to B\pi$, $\bar\Lambda\to \bar B\pi$}
In the helicity frame of $J/\psi\to\Lambda\bar\Lambda\to p\pi^-\bar p\pi^+$, by using the momenta of $p,\pi^-,\bar p$ and $\pi^+$, Eqs.(\ref{jpsiA})(\ref{jpsiB}) can be written as:
\begin{eqnarray}\label{jpsicpa}
A_{J/\psi}=\theta(\sin\theta_1\sin\bar\theta_1\sin(\phi_1-\bar\phi_1)) \nonumber \\
  -\theta(\sin\theta_1\sin\bar\theta_1\sin(\bar\phi_1-\phi_1)),
\end{eqnarray}

\begin{eqnarray}\label{jpsicpb}
B=\sin\theta_1\sin\bar\theta_1\sin(\phi_1-\bar\phi_1),
\end{eqnarray}
With Eq.(\ref{jpsicpa}), one can try to search for the $CP$ violation in $J/\psi\to\Lambda\bar\Lambda$.
Ref.\cite{12pingrg} has shown the result which is consistent with the expectation of $CP$ conservation.

According to Refs.\cite{9bes}\cite{10lambda}, the $\Lambda$ angular distribution parameter $\alpha$ and the product
of $\Lambda$ and $\bar\Lambda$ decay parameters with different final states can be determined by
using the unbinned maximum likelihood method with Eqs.(\ref{ampsim})(\ref{alpham0bar})(\ref{amp0p})(\ref{amp00bar}).
Tab.\ref{paracp} shows the undetermined parameters with different final states. In each channel,
one can get the product of one $\Lambda$ decay parameter and one $\bar\Lambda$ decay parameter.

\begin{table}
\renewcommand\arraystretch{2.0}
\begin{center}
\caption{$\Lambda$ angular distribution parameter
$\alpha$ and the product of $\Lambda$ and $\bar\Lambda$ decay parameters
with different final states}\label{paracp}
\vspace{5mm}
\begin{tabular}{ccc}
Final states & $\Lambda$ angular & $\Lambda,\bar\Lambda$ decay \\
             & distribution      &  parameters  \\
\hline
$p\pi^-\bar p\pi^+$ & $\alpha$ & $\alpha_-\alpha_+$  \\
$p\pi^-\bar n\pi^0$ & $\alpha$ & $\alpha_-\bar\alpha_0$  \\
$n\pi^0\bar p\pi^+$ & $\alpha$ & $\alpha_0\alpha_+$  \\
$n\pi^0\bar n\pi^0$ & $\alpha$ & $\alpha_0\bar\alpha_0$  \\
\hline
\end{tabular}
\end{center}
\end{table}

The advantage of the helicity amplitude analysis for the four channels in Tab.\ref{paracp}
is that four products of $\Lambda$ and $\bar\Lambda$ decay parameters can be determined.
With these four products, $\Lambda$ decay parameters $\alpha_-,\alpha_+$ and $\bar\Lambda$ decay
parameters $\alpha_0,\bar\alpha_0$
can be obtained respectively. BESIII has already accumulated $10^{10}$ $J/\psi$
decays, the detection efficiency
is simulated at least $5\%$ for pure neutral channels and $30\%$ for pure charged
channels with Monte Carlo, the measurements can be done with high precision.
The precision of the measurements is shown in Tab.\ref{precision}.

\begin{table}
\renewcommand\arraystretch{2.0}
\begin{center}
\caption{The precision of the measurements}\label{precision}
\vspace{5mm}
\begin{tabular}{cccc}
Final states & Detection   & $\Lambda,\bar\Lambda$ decay  & Precision  \\
             & efficiency  &  parameters                            &            \\
\hline
$p\pi^-\bar p\pi^+$ & $30\%$ & $\alpha_-\alpha_+$      &  $10^{-3}$ \\
$p\pi^-\bar n\pi^0$ & $15\%$ & $\alpha_-\bar\alpha_0$  &  $10^{-3}$ \\
$n\pi^0\bar p\pi^+$ & $15\%$ & $\alpha_0\alpha_+$      &  $10^{-3}$ \\
$n\pi^0\bar n\pi^0$ & $5\%$ & $\alpha_0\bar\alpha_0$  &  $10^{-2}$ \\
\hline
\end{tabular}
\end{center}
\end{table}

From Tab.\ref{precision}, the precision of four products of $\Lambda$ and $\bar\Lambda$ decay
parameters could be improved by two orders of magnitude compared with the
current values\cite{pdg}. Combining these four products, one can get
$\alpha_-,\alpha_+,\alpha_0$ and $\bar\alpha_0$ respectively. The $CP$
invariance could be obtained by Eq.(\ref{lambdacp}) with $\alpha_-,\alpha_+$ or
$\alpha_0,\bar\alpha_0$, the precision would be also improved.

\section{Summary}
The helicity amplitudes for $J/\psi \rightarrow
 \Lambda \bar{\Lambda}$, $\Lambda(\bar\Lambda)$ decays to different final states
are presented.  Two observables to search for $CP$ violation in
$J/\psi\to\Lambda\bar\Lambda$ can be expressed in the helicity frame
with the information of helicity angles of baryon and antibaryon. Four decay
parameters of $\Lambda$ and $\bar\Lambda$ can be obtained with the joint
helicity amplitude equations by the likelihood fit method which
has been already used in Ref.\cite{9bes}. With the data sample of
$10^{10}$ $J/\psi$ decays accumulated by BESIII, the precision for
$\alpha_-,\alpha_+,\alpha_0$ and $\bar\alpha_0$, compared with
the current value, could be improved
by two orders of magnitude. It would be helpful for the further study
of the $CP$ invariance both in $J/\psi\to\Lambda\bar\Lambda$ and the nonleptonic
decays of $\Lambda$($\bar\Lambda$).

Bin ZHONG, wishes to thank  Rong-Gang PING and Kai
ZHU for stimulating discussions. This work is supported by the National Natural Science Foundation of China under Grant Nos 11305090 and the Research Foundation for Advanced Talents
of Nanjing Normal University (2014102XGQ0085).

\end{document}